\documentclass[%
 reprint, showkeys,groupedaddress,superscriptaddress, amsmath,amssymb,aps]{revtex4-2}

\usepackage{graphicx}
\usepackage{dcolumn}
\usepackage{xcolor}
\usepackage[normalem]{ulem} 
\usepackage{bm}

\begin{document}

\preprint{APS/123-QED}

\title{Bell test of quantum entanglement in attosecond photoionization }

\author{Marco Ruberti}
\email{Corresponding Author \\ e-mail address: m.ruberti11@imperial.ac.uk}
\affiliation{Imperial College London, Physics Department, Blackett Laboratory, Prince Consort Road, SW7 2AZ, London, United Kingdom}
\author{Vitali Averbukh}%
\affiliation{Imperial College London, Physics Department, Blackett Laboratory, Prince Consort Road, SW7 2AZ, London, United Kingdom}
\author{Florian Mintert}%
\affiliation{Imperial College London, Physics Department, Blackett Laboratory, Prince Consort Road, SW7 2AZ, London, United Kingdom}
\affiliation{Helmholtz-Zentrum Dresden-Rossendorf, Bautzner Landstraße 400, 01328 Dresden, Germany}

\newcommand{\mar}[1]{{\color[rgb]{0.6,0,0.6}{#1}}}
\newcommand{\flo}[1]{{\color[rgb]{0,0,0.6}{#1}}}
\newcommand{\vit}[1]{{\color[rgb]{0.6,0,0}{#1}}}
\newcommand{\hil}[1]{{\color[rgb]{0,0.8,0}{#1}}}

\date{\today}

\begin{abstract}
Attosecond physics enables the study of ultrafast coherent electron dynamics in matter upon photoexcitation and photoionization, revealing spectacular effects such as hole migration and coherent Auger dynamics in molecules.
In the photoionization scenario, there has been a strong focus on probing the physical manifestations of the internal quantum coherence within the individual parent ion and photoelectron systems. However, quantum correlations between these two subsystems emerging from the attosecond photoionization event have thus far remained much more elusive.
In this work, we design theoretically and model numerically a direct probe of quantum entanglement in attosecond photoionization in the form of a Bell test. We simulate from first principles a Bell test protocol for the case of noble gas atoms photoionized by ultrashort, circularly polarized infrared laser pulses in the strong-field regime predicting robust violation of the Bell inequality.  
This theoretical result paves the way to the direct observation of entanglement in the context of ultrafast photoionization of many-electron systems.
Our work provides a different perspective on attosecond physics directed towards the detection of quantum correlations between systems born during attosecond photoionization and unravelling the signatures of entanglement in the ultrafast coherent molecular dynamics, including in the chemical decomposition pathways of molecular ions.
\end{abstract}

\keywords{Attosecond Physics, Photoionization, Quantum Entanglement, Bell Test}

\maketitle

\section{\label{sec:lev1}Introduction}

Entanglement is one of the most iconic manifestations of the laws of quantum mechanics. It is a type of correlation between different constituents of a quantum system that has no real analog in classical physics~\cite{Horodecki2009}, and is a key ingredient for quantum information science~\cite{Nielsen2010}, where it can be used as a resource for quantum computation, metrology and imaging~\cite{Horodecki2009,Nielsen2010,Steinlechner2013,Degen2017,Bongs2019,Defienne2021}. 
Quantum entanglement has also received much attention in the context of atomic and molecular physics, with particular focus on the properties of the emitted entangled photons~\cite{Radtke2008,Fratini2011,Tichy2011}. However, the role played by quantum entanglement in the electron dynamics in atoms and molecules remains largely unexplored and its effect on the physical processes occurring on the natural electronic, i.e. attosecond, timescale is yet to be understood.

The main goal of attosecond physics is to resolve ultrafast electron dynamics in a wide variety of atomic, molecular, and condensed phase systems~\cite{Krausz2009}. Following the recent spectacular advances in the synthesis and characterization of ultrashort laser pulses~\cite{Leone2014,Calegari2016}, of time duration ranging from few tens of femtoseconds down to hundreds of attoseconds, a large variety of spectroscopic and imaging techniques reaching attosecond time resolution have been developed including methods based on 
strong-field light-matter interaction such as high-order harmonic generation (HHG) spectroscopy~\cite{Itatani2004}, laser-induced electron diffraction~\cite{Niikura2002}, photoelectron holography~\cite{Huismans2011} attosecond streaking~\cite{Itatani2002} and attosecond pump-probe spectroscopy~\cite{Calegari2014,Barillot2021,Schwickert2022}. The key physical process that lies in the basis of these 
techniques, whether relying on HHG sources~\cite{Johnson2018} or on X-ray free electron lasers~\cite{Duris2020}, is photoionization. While photoionization is certainly one of the universally acknowledged manifestations of the quantum nature of matter that led, for example, to the discovery of energy quanta, it is far less widely appreciated that it also bears the mark of quantum entanglement~\cite{Akoury2007,Schoffler2008,Goulielmakis2010,Ruberti2019PCCP,Vrakking2021,Koll2022}. 

During the photoionization process, an originally bound many-electron quantum system, such as an atom or a molecule, is broken up, in the simplest scenario, into two subsystems: the emitted photoelectron and its parent ion.
In the cases where, for example, the parent ion can be found in multiple quantum states, one can expect such multi-channel photoionization to result in an entangled state of the photoelectron and the ion, in which neither of the produced spatially separated particles can be assigned particular values of the physical observables, such as energy, angular momentum or spin. The most immediate consequence of the presence of entanglement between the photoelectron and the parent ion upon photoionization is that each subsystem is in general characterized by a mixed quantum state possessing a reduced degree of internal quantum coherence~\cite{Goulielmakis2010,Ruberti2021,Ruberti2022,Alexander2023}. 
A pioneering work was able to demonstrate the reduced coherence of the ionic system by predicting theoretically and characterizing experimentally
the density matrix of atomic krypton ions emerging from strong field ionization~\cite{Goulielmakis2010}, while the experimental reconstruction of the photoelectron's density matrix upon atomic photoionization has also been reported more recently~\cite{Bourassin2020}.

However, the emerging picture of the interplay of coherence and entanglement in photoionization has so far not been supported by any quantitative measurement of the quantum entanglement, and no procedure for such a measurement has been proposed. Indeed, the common description of ultrafast photoionization experiments has concentrated on the internal, ultrafast dynamics that is underpinned by the quantum coherence within each of the subsystem produced by photoionization, such as hole migration~\cite{Schwickert2022} in the ion, while the quantum correlations between the photoelectron and the parent ion have remained elusive in attosecond physics. 

Moreover, the majority of studies of photoionization and the ensuing electron processes have treated the parent ion and the photoelectron as if they were individual, isolated systems, i.e. fully neglecting their mutual quantum correlations. One example of such an approach is provided by the bulk of theoretical work on hole migration following photoionization of polyatomic molecules, including small biomolecules~\cite{Breidbach2003,Kuleff2014,Despre2015,Kuleff2016,Du2019,Matselyukh2022,Gu2023}. In these works, the so-called sudden approximation is invoked to approximate, in the limit of large photon energy, the state of the parent ion as a pure, coherent superposition of energy eigenstates, which is predicted to feature a migration of the positive charge across the backbone of the molecular ion on a sub- to few- femtosecond timescale~\cite{Breidbach2003}. The sudden approximation picture explicitly neglects the effect of the entanglement between the photoelectron and the molecular cation. In a similar way, majority of the studies aiming to characterize photo- and secondary (e.g. Auger) electrons neglected the entanglement between the 
emitted photoelectron and the parent ion or between the photoelectron and Auger electron in the post-collision interaction scenario, where the energy exchange between the two consecutively emitted particles has been treated semi-classically~\cite{niehaus1977,eberhardt1988,guillemin2015}. 

Nevertheless, already in an early work~\cite{Schmidtke2000} it was realized that measurement of the entire set of observables for the photoelectron only is not sufficient to obtain a complete characterization of an inner-shell photoionization process~\cite{Schmidtke2000}, and that additional information about the atomic cation is needed~\cite{Snell2001}. This was clearly indicating the presence of correlations between the two subsystems, even though their nature, classical or quantum, was not revealed then yet. Moreover, starting from the early 2000s, a few studies have explored the role of entanglement in attosecond processes. Some 
of these works have focused on entanglement between the photoelectron and the parent ion~\cite{Fedorov2004,Chandra2004,Kim2004,Spanner2007,Spanner2007b,Vatasescu2013,Czirjak2013,Chakraborty2015,Majorosi2017,Nishi2019,Ruberti2019PCCP,Ruberti2021,Alexander2023,Vrakking2021,Koll2022,Vrakking2022b,Nandi2023}, 
while other studies have focused on electron-electron entanglement~\cite{Liu1999,Christov2019,Omiste2019,Maxwell2022}. 
A pioneering experimental work~\cite{Schoffler2008} has been able to probe hole localization after inner-shell ionization of molecular nitrogen by measuring in coincidence the ultrafast Auger electron angular emission patterns and the spectrum of nuclear fragments. The results showed that observation of symmetry breaking (electronic hole localization), or preservation (electronic hole delocalization) depends on how the quantum entangled Bell state created by Auger decay is detected by the measurement. More recently, experimental investigations of ultrafast preparation of atomic Bell-like states~\cite{Eckart2022}, as well as of ultrafast control mechanisms for vibrational and electronic entanglement~\cite{Vrakking2021,Koll2022,Shobeiry2022} have also been performed. However, these studies have investigated photoelectron-ion entanglement in an indirect way, mostly by monitoring the change of  internal quantum coherence in one of the entangled subsystems~\cite{Vrakking2021,Koll2022,Vrakking2022b,Nishi2019}. The direct test for quantum entanglement is given by verification of the violation of Bell inequalities~\cite{Bell1964,Kocher1967,Pan2000,Lee2011,Hensen2015}.
If applied to photoionization, the Bell test should be, in its simplest realization, based on detecting correlations in the measurements of two noncommuting observables for both the parent ion and the photoelectron. Such a direct and rigorous verification and quantification of the quantum entanglement produced upon photoionization is still missing and the direct experimental evidence of this type of entanglement has so far been considered to be ``often practically impossible, requiring the measurement of incompatible observables, such as momentum and position"~\cite{Maxwell2022}. 

In this work, we conceptually design and numerically simulate a Bell-test experiment for the case of atomic photoionization. Specifically, we consider argon atoms photoionized by ultrashort, circularly polarized infrared (IR) laser pulses in the strong-field regime. We derive Bell inequalities which are sensitive to the entanglement between the internal energy eigenstates of the atomic ion and the spin states of the free photoelectron. We show that by exploiting the spin polarization of the photoelectron beam, one can observe a robust violation of the Bell inequalities, demonstrating quantum entanglement between the photoelectron and the atomic ion. Our results demonstrate a quantum protocol for rigorous detection of quantum entanglement in ultrafast photoionization and pave the way to the direct observation of entanglement in the context of attosecond physics.

\section{\label{sec:lev2}Protocol for designing a Bell test}

\subsection{\label{subsec:A1}Choice of Bell inequality type and photoelectron observables}

In order to verify and quantify the entanglement between the photoelectron and its parent ion, it is necessary to perform coincidence measurements on the two subsystems. In this type of measurement set-ups, the quantum correlations are reflected by the coincidence statistics, which are sensitive to which observables of the emitted photoelectron and its parent ion are measured. 

In this work, we construct Bell inequalities of the Clauser, Horne, Shimony, Holt (CHSH) type~\cite{Clauser1969}, which require the measurement of two different noncommuting observables per subsystem. This type of Bell inequalities is also based on dichotomic observables, i.e. on observables adopting only two values, which can be arbitrarily set equal to $\pm 1$. 
The most common Bell test for a pair of entangled particles requires the evaluation, in separate experiments, of the quantity $\mathbb{S}$ combining four quantum correlations functions of the particles pair in the following way,
\begin{eqnarray}\label{eq:BellInequalityGEN}
     \mathbb{S} \,=\,   \langle\hat{A}\left(a\right)\otimes\hat{B}\left(b\right)\rangle\,-\,\langle\hat{A}\left(a\right)\otimes\hat{B}(b^{'})\rangle\,+\, \nonumber \\ +\,\langle\hat{A}(a^{'})\otimes\hat{B}\left(b\right)\rangle\,+\,\langle\hat{A}(a^{'})\otimes\hat{B}(b^{'})\rangle\,,
\end{eqnarray}
where $a,a^{'}$ and $b,b^{'}$ denote different settings for the detector of particle A and of particle B, respectively. 
Violations of the CHSH inequality, i.e. $\mathbb{S} > 2$, are predicted for entangled states by the laws of quantum mechanics, which also set a maximum value for the value of $\mathbb{S}$ of $2\sqrt{2}$, known as Tsirelson's bound~\cite{Cirel1980}. 

A key step to design a Bell test of the CHSH type consists of correctly identifying the appropriate pairs of noncommuting observables to be measured for each subsystem. In the case of a free photoelectron resulting from a photoionization event, the set of observables that could in principle be measured, and that also give a complete description of the photoelectron's state, consists of its kinetic energy $\xi$, its linear orbital momentum (angle of emission) or orbital angular momentum, and its spin angular momentum. As a result of photoionization, the entanglement between the photoelectron and the cation can in principle be encoded in each of these different degrees of freedom of the photoelectron. 

Among all the possible corresponding observables, the kinetic energy is not a good one for entanglement tests because it commutes with all the other photoelectron observables. On the contrary, Bell inequalities can in principle be constructed using different, noncommuting components of either the orbital or the spin angular momentum of the free photoelectron. In general, components of the two angular momenta, as well as noncommuting components of linear and angular momenta of the free photoelectron, are also potentially good candidates to construct a Bell test. 

Despite the recent progress in the measurement of the orbital angular momentum of electron beams as demonstrated in~\cite{Grillo2017a,Grillo2017b,Larocque2018},
this observable remains, in general, difficult to measure with high accuracy and its detection is still affected by considerable experimental error~\cite{Grillo2017a,Grillo2017b,Larocque2018}. On the other hand, examples of accurate measurements of the projection of a free electron's spin along a specific direction in space are available in the literature. 
The spin of the photoelectron can indeed be measured using a Mott detector~\cite{Hartung2016,Trabert2018}, which relies on spin–orbit interactions and the consequent spin-dependent asymmetry in the scattering of electrons at high-Z atoms. Retarding-potential Mott polarimeters have been used to perform single-particle-resolved measurements of electron's spin and spin coincidence measurements of scattered electron pairs~\cite{Berezov2010}.
Here it is important to note that the overall efficiency of an experiment based on coincidence detection will also depend on the efficiency of the Mott analyzer used for spin detection. In the literature optimized efficiencies are reported to be approximately $2 \times 10^{-4}$~\cite{Burnett1994}. Alternative measurements of the photoelectron spin have also been theoretically simulated in~\cite{Batelaan1997,Garraway1999,McGregor2011} and they are based on the use Stern–Gerlach-like devices.

\subsection{\label{subsec:A2}The target system and photoionization regime: the case of noble gas atoms photoionized by circularly polarized light}

Following the considerations of Sec.~\ref{subsec:A1}, in this work we will explore an entanglement test based on the observation of the spin angular momentum of the photoelectron. This requires quantum entanglement to be encoded in the spin degree of freedom of the photoelectron. The aforementioned condition can be achieved, for example, by photoionization of noble gas atoms with intense, circularly polarized IR laser pulses under standard experimental conditions. Indeed, it was both predicted theoretically~\cite{Barth2013,Barth2014}, and verified experimentally~\cite{Herath2012,Hartung2016,Liu2018,Trabert2018}, that this type of ionization process gives rise, in energy-resolved measurements, to photoelectron beams characterized by a high degree of spin polarization. Previous analytical calculations also predicted the polarization of the photoelectron beam to depend on the specific ionization channel~\cite{Barth2013,Trabert2018}, which reflects the entanglement between the internal spin-orbit states of the parent ion and the projection of the emitted electron's spin along the propagation direction of the ionizing laser beam. 
Spin polarization arises due to the interplay of the electron-ion entanglement and the strong dependence, in intense circularly polarized fields, of the ionization probability to the sense of electron rotation in the initial state, i.e. the sign of the magnetic quantum number $m_l$ of the orbital the electron is removed from. 
In the case of right-handed circularly polarized fields, ionization from a co-rotating orbital ($p_{+1}$) is much lower than the one from a counter-rotating orbital ($p_{-1}$). In particular, low-energy photoelectron emission has been predicted to be strongly spin-polarized, due to the suppression of emission of corotating ($p_{+1}$) electrons~\cite{Barth2013,Barth2014}. 

Here it is important to note that, even though in the present work we focus our attention on the case of noble gas atoms strong-field ionized by circularly polarized pulses, the entanglement protocol that we construct is by no means restricted to such systems. Other systems where quantum entanglement can be encoded in the spin degree of freedom of the spin-polarized photoelectron include, for example, linear and ring-shaped molecules with a degenerate highest-occupied-molecular-orbital (HOMO) and a singlet ground state~\cite{Barth2013}. In such systems spin-orbit states have lower degeneracy than in atoms and higher degree of total integrated spin polarization are expected. Other photoionization regimes that feature in spin polarization of the photoelectron beam include single-photon ionization, either from a particular fine structure level of an atom or a molecule~\cite{Cherepkov1981} or in the vicinity of the Cooper minima in the photoionization continua~\cite{Fano1969}, and resonant multi-photon ionization in the perturbative limit~\cite{Lambropoulos1973,Dixit1981}. Importantly, a high degree of spin polarization (close to $100\%$) is not always associated with minima in cross sections, but can be achieved, e.g., away from the minimum in the three-photon ionization cross section of alkali-metal atoms~\cite{Teague1976} and at the maximum of the one-photon cross section for Xe~\cite{Nakajima2002}.

\begin{figure*}
\includegraphics[scale=0.8]{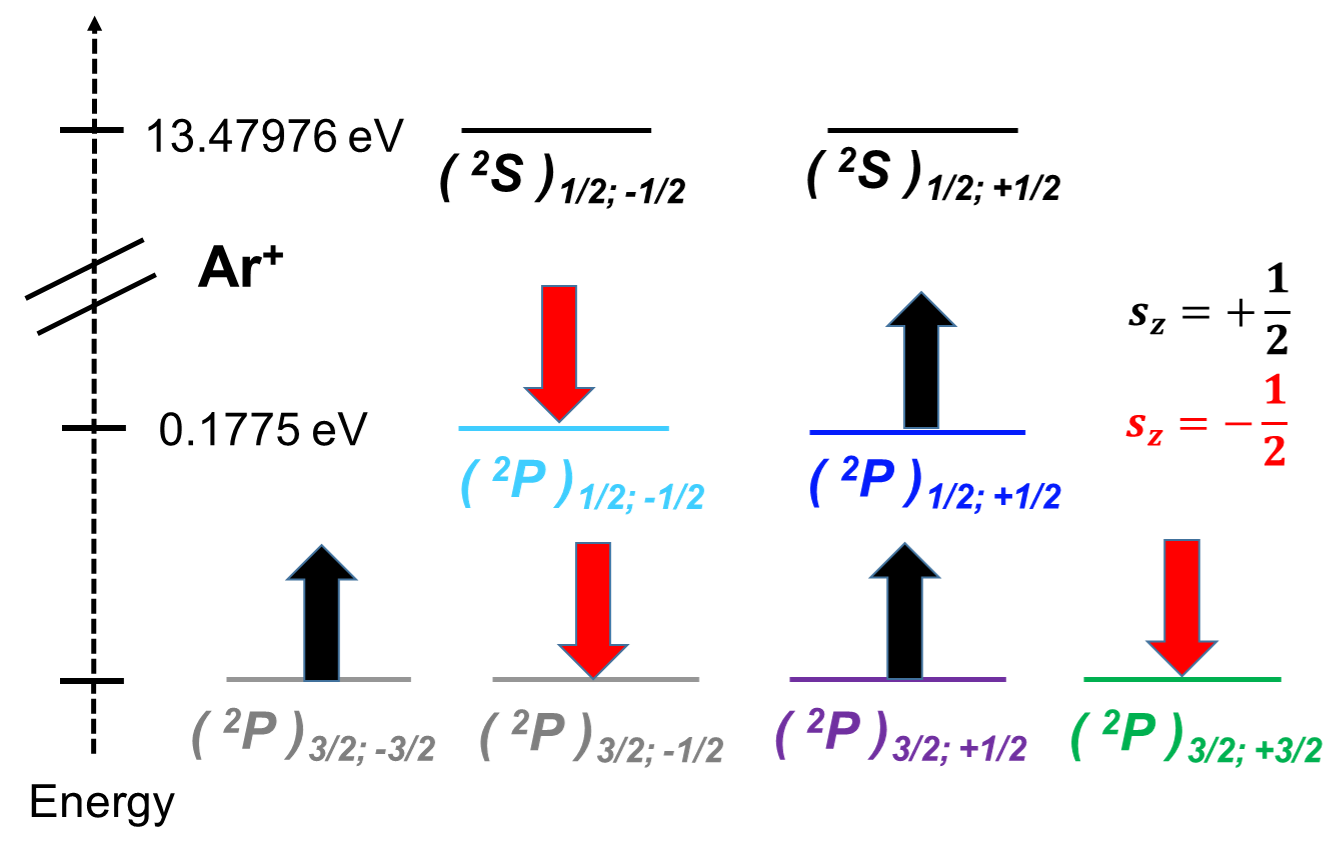}
\caption{\label{fig:Levels} Energy levels of Ar$^{+}$ parent ion. In the bottom part of the figure, the 4-fold degenerate ground state $^{2}P_{3/2}$, and the two-fold degenerate first excited state $^{2}P_{1/2}$, which are populated upon strong-field ionization by circularly polarized IR laser pulses, are shown. The specific spin orientation, along the propagation direction of the ionizing laser pulse, of the correspondent photoelectron is also shown (for the case of right-handed laser polarization). The second, higher-lying, excited state $^{2}S_{1/2}$ is also shown.}
\end{figure*}

In the theoretical scheme presented here, the many-electron system considered consists of a neutral N-electron atom that is initially, i.e. before photoionization, in its electronic ground state $|\Psi_{N}^{0}\rangle$. The states of the parent ion that are populated by the photoionization process, and that are relevant for the measurement protocol proposed here, are shown in Fig.~\ref{fig:Levels}. The yield of different ionization channels in strong-field ionization (SFI) depends exponentially on their respective ionization potential. Therefore, in the case of Ar$^{+}$ only the two lowest-energy levels $^{2}P_{3/2}$ and $^{2}P_{1/2}$ are effectively populated upon photoionization. The spin-orbit splitting of these two energy levels is $\Delta E = 0.1775$ eV, with an ionization potential of the parent ion ground state $^{2}P_{3/2}$ equal to $I_{P} = 15.75962$ eV. The ionization potential of the second excited state $^{2}S_{1/2}$ of Ar$^{+}$ is much higher ($29.23938$ eV) and therefore the population of this state upon SFI can be safely neglected. 
SFI of $p_0$ electrons by circularly polarized light is strongly suppressed and will be neglected here. Both approximations are fully justified based on previous analytical calculations~\cite{Barth2013}, and the high accuracy of this approximation has been confirmed by the results of the \textit{ab initio} simulations performed in this work, as we will show in Sec.~\ref{subsec:results}. 

As a result of photoionization dynamics and of the coupling of the spin and orbital angular momenta as given by the Clebsh-Gordan coefficients, a single specific spin state of the photoelectron corresponds to each ionization channel. This is illustrated in Fig.~\ref{fig:Levels} for the case of right-handed circular polarization (for opposite polarization the spin components will also be inverted). 

Thus, the complete photoionized state can be written as
\begin{eqnarray}
|\Psi^{F}_{N}\rangle =  \sum_{j=1}^{6} \iint d\xi d\alpha\
C_{js_j}\left(\xi,\alpha \right)\ |^{2}P_{l_jm_j};\, \xi \, s_j \, \alpha \rangle\ ,
\label{eq:statetot}
\end{eqnarray}
$s_1=s_3=s_6=\,\uparrow$, $s_2=s_4=s_5=\,\downarrow$,
$l_1=l_4=1/2$, $l_2=l_3=l_5=l_6=3/2$, $m_2=\,3/2$, $m_1=m_3=\,1/2$, $m_4=m_5=\,-1/2$, $m_6=\,-3/2$, where the photoelectron is described by its kinetic energy $\xi$, the sign of its spin component along the propagation direction of the photoionizing pulse, $\bf{\uparrow/\downarrow}$, and any extra degree of freedom denoted by $\alpha$, e.g. $\alpha\,=\,\left(l,\,m \right)$ or 
$\alpha\,=\,\left(\theta,\,\phi \right)$.

The spin polarization $Pol_{S}$ of the photoelectron emitted from the $^{2}P_{3/2}$ channel is defined as
\begin{equation}
Pol_{S}^{^{2}P_{3/2}} = 
\frac{\sum_{j}^{3,6} 
|C_{j\uparrow}\left(\xi,\alpha \right)|^2
\,-\,\sum_{j}^{2,5} 
|C_{j\downarrow}\left(\xi,\alpha \right)|^2}{\sum_{j}^{2,3,5,6} 
|C_{js_j}\left(\xi,\alpha \right)|^2}
\label{eq:polarization}
\end{equation}
Analogous definition can be written down for the $^{2}P_{1/2}$ channel. The maximum value of the channel-resolved spin polarization along the propagation direction of the photoionizing laser pulse is $-50\%$ and $100\%$ for the $^{2}P_{3/2}$ and the $^{2}P_{1/2}$ ionization channels, respectively. In order to approach such a strong spin polarization, it is necessary to suppress ionization from the $p_{+1}$ atomic orbital, which populates the ionic states with negative $M_{J}$, and mainly have emission of electrons from the $p_{-1}$ atomic orbital, which leads to the population of the $^{2}P_{1/2,+1/2}$, $^{2}P_{3/2,+3/2}$ and $^{2}P_{3/2,+1/2}$ states of the parent ion. This can be achieved by using right-handed circularly polarized pulses. 

Moreover, in order to obtain a significant level of entanglement in the photoionized state, it is necessary for the ionization yields corresponding to the two energy channels of the parent ion, separated in energy by the ground-state spin-orbit splitting, to have comparable values. As it will be shown in the next section, this requirement will be fully satisfied in the case of Ar$^{+}$.

\begin{figure*}
\includegraphics[scale=0.8]{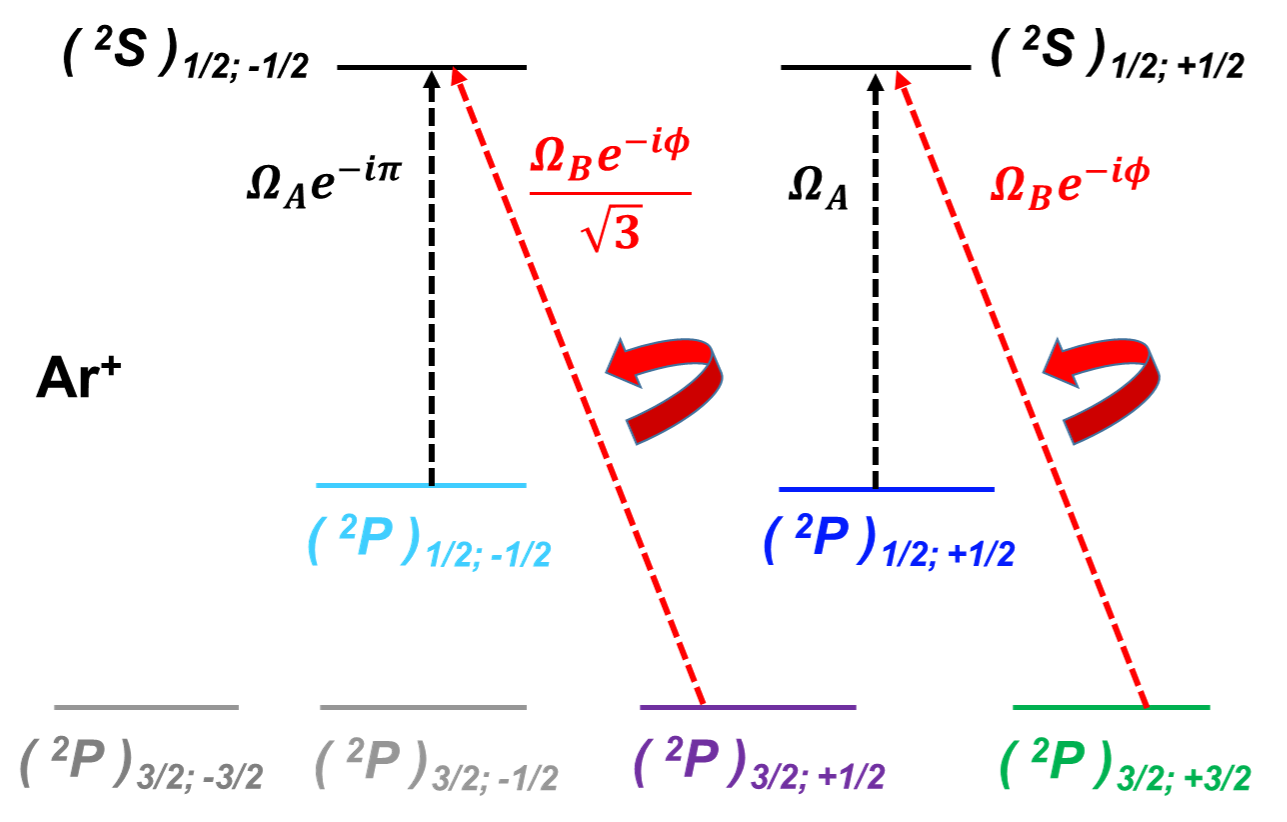}
\caption{\label{fig:Scheme}Schematic representation of the laser pulses to be applied to the Ar$^{+}$ parent ion subsystem within the entanglement detection protocol. Resonant, circularly polarized (red dashed lines) and linearly polarized (black dashed lines) laser light couple, respectively, the initially-populated ground $^{2}P_{3/2}$ and first excited $^{2}P_{1/2}$ state to the, initially not populated, second excited state $^{2}S_{1/2}$. 
The corresponding coupling matrix elements are also shown; $\Omega_{a}$ and $\Omega_{b}$ are the Rabi frequencies and $\phi$ is the relative phase between the two electric fields.}
\end{figure*}

\subsection{\label{subsec:A3}Choice of observables for the cation and formulation of the Bell inequality}

Finally, in order to design a Bell test, it is necessary to identify the choice of the pair of ionic operators, $\hat{Q}^{\text{Ion}}$ and $\hat{R}^{\text{Ion}}$, that will result in the clearest verification of entanglement, i.e. in the strongest violation of the Bell inequality, for the states resulting from photoionization.
Direct, projective measurements on the parent ion can be concretely realized in the basis of its energy eigenstates. The combination of this type of measurement with further coherent operations, to also be performed on the parent ion, allows one to construct generic ionic operators $\hat{Q}^{\text{Ion}}$ and $\hat{R}^{\text{Ion}}$. The states of the parent ion, which encode entanglement with the photoelectron, are characterized by the same (P) orbital symmetry. As a result, dipole transitions between them are symmetry-forbidden by selection rules. Coherent operations can nevertheless be performed by increasing the active Hilbert space to the second 
excited state of S symmetry.
The pulse scheme, and the corresponding Hamiltonian matrix elements between the states involved, are schematically illustrated in Fig.~\ref{fig:Scheme}.

We propose to use two synchronized laser pulses, delayed with respect to the IR ionizing one: the first pulse is characterized by circular polarization, co-propagates in the same direction as the IR ionizing pulse, and its photon energy is in resonance with the ionic transition between the $^{2}P_{3/2}$ ground state and the $^{2}S_{1/2}$ second-excited state energy levels; the second one has linear polarization along the propagation direction of the first pulse, and its resonant with respect to the transition between the $^{2}P_{1/2}$ first-excited state and the $^{2}S_{1/2}$ second-excited state energy levels.
The first, circularly polarized pulse couples the $^{2}P_{3/2,+3/2}$ and $^{2}P_{3/2,+1/2}$ states to the $^{2}S_{1/2,+1/2}$ and $^{2}S_{1/2,-1/2}$ states, respectively. The Rabi frequencies of the two transitions differ by a factor of $\frac{1}{\sqrt{3}}$, originating from the Clebsh-Gordan coefficients describing the $\hat{L}-\hat{S}$ coupling of the $\hat{J}^{2},\,\hat{J}_{z}$ eigenstates considered. 
The linearly polarized pulse couples the $^{2}P_{1/2,+1/2}$ and $^{2}P_{1/2,-1/2}$ states to the $^{2}S_{1/2,+1/2}$ and $^{2}S_{1/2,-1/2}$ states, respectively. The matrix elements corresponding to the two transitions differ by a $\pi$ phase factor, again originating from the $\hat{L}-\hat{S}$ coupling Clebsh-Gordan coefficients. 
The two pulses are applied to the parent ion subsystem for a time \textit{T}. 

As it can be seen in Fig.~\ref{fig:Scheme}, this choice of pulses gives rise to two independent 3-level $\Lambda$ systems, thus allowing one to effectively couple to each other the $^{2}P_{1/2,+1/2}$ and $^{2}P_{3/2,+3/2}$ ionic states, as well as $^{2}P_{1/2,-1/2}$ and $^{2}P_{3/2,+1/2}$, by using an extra state. In addition, the designed coherent operations do not affect the $^{2}P_{3/2,-1/2}$ and $^{2}P_{3/2,-3/2}$ states of the parent ion, whose populations remain constant in time and equal to their value upon SFI by the IR ionizing pulse. On the contrary, some of the population that was initially prepared by the SFI is transferred by the applied pulses to the $^{2}S_{1/2}$ excited energy level. 

In order to complete the design of the ionic operators to be used in the construction of the Bell inequality, we propose a dichotomic projective measurement $\hat{O}^{\text{Ion}}$ that distinguishes whether the parent ion has (measurement result +1) or has not (measurement result -1) been excited to the $^{2}S_{1/2}$ energy level after the application of our pulse scheme: 
\begin{eqnarray}
\hat{O}^{\text{Ion}}\,=\,+1\,\sum_{M_{J}}|^{2}S_{1/2,M_{J}}\rangle \langle ^{2}S_{1/2,M_{J}}| \nonumber \\
\,-1\,\sum_{M_{J}}|^{2}P_{3/2,M_{J}}\rangle \langle ^{2}P_{3/2,M_{J}}| \nonumber \\
   \,-1\,\sum_{M_{J}}|^{2}P_{1/2,M_{J}}\rangle \langle ^{2}P_{1/2,M_{J}}|.
\end{eqnarray}
This measurement can be implemented in the laboratory by, e.g., observing the fluorescence emission resulting from the population of the $^{2}S_{1/2}$ state. 
    
In practice, the application of the two laser pulses allows one to realize dark and bright states as linear combinations of the two, effectively coupled, initially-populated states. 
The general form of the effective ionic operator (a 4x4 matrix on the space spanned by ionic states 
$|^{2}P_{3/2,+3/2}\rangle$, 
$|^{2}P_{3/2,+1/2}\rangle$, 
$|^{2}P_{1/2,+1/2}\rangle$,
and $|^{2}P_{1/2,-1/2}\rangle$) that can be obtained with this procedure is detailed in App. B.
In particular, the matrix elements of the effective ionic operator have a parametric dependence on the following parameters, $E_{\text{Lin}},\,E_{\text{Cir}},\,\phi,\,\textit{T}$, of the two laser pulses used. Different dark and bright states can be obtained by varying these parameters. This in turn allows one to construct different, noncommuting ionic operators $\hat{Q}^{\text{Ion}}$ and $\hat{R}^{\text{Ion}}$ (see App.~\ref{sec:app2} for the derivation and the form of these operators).

As a result of their transparency to the applied pulses, the $^{2}P_{3/2,-1/2}$ and $^{2}P_{3/2,-3/2}$ states of the parent ion do not contribute to the Bell inequality. 
With the laser parameters used in this work, the initial population of these states upon SFI is much smaller than the one of the other ionic states; therefore, as we will see in Sec.~\ref{sec:lev3},  this does not compromise the successful detection of entanglement. 

Here it is also worth noting that the frequency of the resonant laser pulses, which need to be applied to the parent ion in order to perform these coherent operations, is sufficiently off resonance not to couple the $|^{2}S_{1/2}\rangle$ excited state to doubly-ionized states at or above the double ionization potential of the system.

Combining the measurements on the parent ion subsystem with the two independent spin measurements on the photoelectron, as given by the operators $2\hat{S}_{z}$ and $2\hat{S}_{x}$, the violation of the resulting Bell inequality can be written as (see App.~\ref{sec:app2} for the derivation in terms of the photoionized state coefficients):  
\begin{eqnarray}\label{eq:BellInequality}
  \mathbb{S} \,=\, \langle \hat{Q}^{\text{Ion}}\otimes2\hat{S}_{x}\rangle\,-\,\langle\hat{Q}^{\text{Ion}}\otimes2\hat{S}_{z}\rangle\,+\, \nonumber \\ +\,\langle\hat{R}^{\text{Ion}}\otimes2\hat{S}_{x}\rangle\,+\,\langle\hat{R}^{\text{Ion}}\otimes2\hat{S}_{z}\rangle\,\geq 2  ~.
\end{eqnarray}

In order to estimate the value of $\mathbb{S}$, as given by Eq.~\ref{eq:Bell}, and the violation of the Bell inequality, we need to accurately calculate the photoionized state of the bipartite photoelectron and parent ion state (Eq.~\ref{eq:state1}) produced upon SFI or neutral argon, and to select the values of the laser pulses parameters.

\section{\label{sec:lev3}Violation of the Bell inequality: results by \textit{ab initio} simulations}

\subsection{\label{subsec:ADC} \textit{Ab initio} method and numerical parameters}

The accurate, from-first-principles, prediction of the state of the full bipartite system that is produced upon multi-channel photoionization on a many-electron system, and the simulation of a Bell test experiment, require the use of advanced methodologies for photoionization dynamics that can describe, on an equal footing, the bound interacting many-electron system and the photoelectron states. In our work, we apply the 
advanced, \textit{ab initio} time-dependent B-spline algebraic diagrammatic construction (ADC) method for many-electron photoionization dynamics~\cite{Ruberti2014,Averbukh2018,Ruberti2019,Ruberti2023} to compute the state coefficients of Eq.~\ref{eq:statetot}, as well as the populations and coherences of Eq.~\ref{eq:initial_matrix}, resulting from strong field ionization of argon atom. 

In this work, the TD B-spline ADC method, already extensively applied to the study of attosecond many-electron dynamics in atoms and molecules~\cite{Ruberti2014,Averbukh2018,Ruberti2018,Ruberti2019,Ruberti2019PCCP,You2019,Ruberti2021,Ruberti2023}, was extended by us to describe the effect of spin-orbit couplings. In particular, we have used the TD B-spline ADC approach at first level of theory (ADC(1)) in the ADC(n) hierarchy; we have solved the N-electron time-dependent Schr{\"{o}}dinger equation for the argon atom interacting with the intense IR laser field, given by
\begin{equation}
    \textit{i}\hbar \frac{\partial |\Psi^{N}\left(t\right)\rangle}{\partial t} \,=\, \left[\hat{H}_{0}+\hat{D}\cdot\mathbf{E}\left(t\right)\right]|\Psi^{N}\left(t\right)\rangle
    \label{eq:method1}
\end{equation}
by expressing the time-dependent state of the composite N-electron system as a linear combination of the neutral argon ground state and the manifold of one-hole–one-particle (1h-1p) excitations of the latter, 
\begin{equation}
    |\Psi^{N}\left(t\right)\rangle\,=\,
    C_{0}\left(t\right)|\Psi^{N}_{0}\rangle\,+\,
\sum_{i,a}C_{i,a}\left(t\right)|\Psi^{N}_{i,a}\rangle ~.
\label{eq:method2}
\end{equation}
Here the hole and particle indices, corresponding to the occupied and virtual Hartree-Fock orbitals of the neutral system, are indicated by \textit{i} and \textit{a}, respectively.
In Eq.~\ref{eq:method1} $\hat{H}_{0}$ is the field-free N-electron ADC Hamiltonian, $\hat{D}$ is the electric dipole operator and $\mathbf{E}\left(t\right)$ the time-dependent electric field vector of the ionizing laser pulse. 

In order to describe the effect of the spin-orbit couplings, we have worked with 1h-1p configurations (as well as energy eigenstates) 
that are eigenfunctions of the total angular momentum $\hat{J}=\hat{L}+\hat{S}$, i.e. $|\Psi^{N}_{i,a}\rangle = |\Psi^{N}_{i,a} \left(J,M_{J},L,S\right)\rangle$, and we have replaced the nonrelativistic Hartree-Fock energies of the hole orbitals $\psi^{\left(j,m_{j},l,s\right)}_{i}$ with the experimental values of the ionization potentials of argon atom.  

The B-spline ADC method at the ADC(1) level of theory has been successfully used in the strong-field regime to model the intensity-dependent interference minimum that is present in the HHG spectra of the CO$_{2}$ molecule~\cite{Ruberti2018HHG}. 
Inclusion of the 1h-1p manifold in the photoionization dynamics, allows one to describe excitation of the neutral argon atom in any of the singly-excited bound states as well as its photoionization into a bipartite, photoelectron \& Ar$^{+}$, system. The explicit and accurate representation of the photoelectron's wavefunction is achieved by means of the B-spline single-particle basis set~\cite{Ruberti2014}. The time propagation of the unknown coefficients $C_{i,a}\left(t\right)$ of the N-electron state is performed by means of the Arnoldi–Lanczos algorithm~\cite{Ruberti2018HHG}. We calculated the channel-resolved, energy-dependent photoelectron angular distributions and spectra of Eq.~\ref{eq:densityDIAG} using the time-dependent surface flux technique~\cite{Bray2021}. 

In the simulation, we used a parabolic-linear sequence for the B-spline knots~\cite{Ruberti2014}, a radial box radius of $R_{max} = 1100$ atomic units and a total of $N_{b} = 1300$ radial B-spline functions. The maximum angular momentum employed in the expansion of the angular part of the photoelectron wavefunction in spherical harmonics was $l_{max} = 20$. We verified that all the results were fully converged with respect to our choice of the basis set parameters. 
In order to absorb the wavefunction and avoid its reflections from the grid boundary, we also included in the simulation a complex absorbing potential with starting radius $R_{abs} = 730$ atomic units.

In order to have a strong violation of the Bell inequality, the ionic states in the two $^{2}P_{3/2}$ and $^{2}P_{1/2}$ energy levels need to be populated upon photoionization as equally as possible. The relative population of the two energy levels, separated by an energy gap determined by the spin-orbit splitting of the nonrelativistic ground state, upon ionization strongly depends on the ionizing laser parameters (intensity and frequency) used. This dependence helps one to identify the optimal ranges of laser intensity and frequency that can potentially give rise to the strongest violation. 
In this work, we have used an ultrashort bandwidth-limited circularly polarized laser pulse characterized by a peak intensity equal to $I_{peak} = 5 \times 10^{13} W/cm^{2}$ and a central photon energy $\hbar\omega_{central} = 2$ eV, which corresponds to a $\lambda = 620$ nm central wavelength. The vector potential, in the x-y plane, of the right-handed circularly polarized pulse is described by a cosine squared envelope 
\begin{equation}
\mathbf{A}\left(t\right)=  
A_{0}\cos^{2}\left(\pi \frac{t}{\tau}\right) \cdot \left(\cos\left(\omega t \right),\sin\left(\omega t \right),0\right) ~. 
\end{equation}
The total pulse duration $\tau$ is set in our simulations to 3 laser cycles in terms of $\omega_{central}$, i.e. $\tau \approx 6$ femtoseconds.

\begin{figure*}
\includegraphics[scale=1.1]{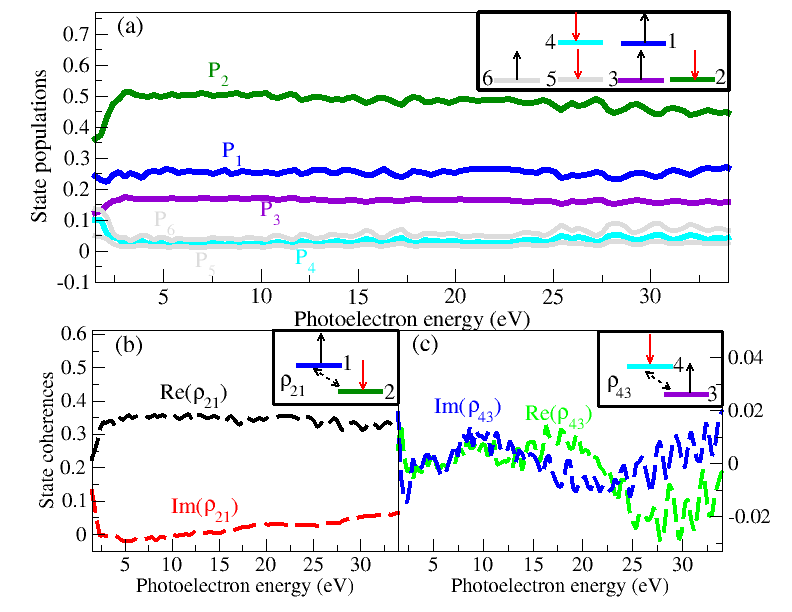}
\caption{Panel (a) -
Normalized, angle-integrated, kinetic-energy dependent diagonal elements of the effective density matrix of Eq.~\ref{eq:densityDIAG} upon photoionization of Ar by a circularly polarized 3-cycle laser pulse with peak intensity $I_{peak} = 5 \times 10^{13} W/cm^{2}$ and $\hbar\omega_{central}= 2$ eV central photon energy. The diagonal matrix elements $\tilde{\rho}^{\xi}_{i,i}=P_{i}$ correspond to the populations $P_{i} $ of the 6 states of Eq.~\ref{eq:statetot} and Figs~(\ref{fig:Levels},\ref{fig:Scheme}). 
Panel (b) - Normalized, angle-integrated, kinetic-energy dependent $\tilde{\rho}^{\xi}_{21}$ off-diagonal element of the effective density matrix (see Eq.~\ref{eq:densityOFFDIAG}) between the states $|^{2}P_{3/2,+3/2};\downarrow\rangle$ and $|^{2}P_{1/2,+1/2};\uparrow\rangle$. Panel (c) - Normalized, angle-integrated, kinetic-energy dependent $\tilde{\rho}^{\xi}_{43}$ off-diagonal element of the effective density matrix (see Eq.~\ref{eq:densityOFFDIAG}) between the states $|^{2}P_{1/2,-1/2};\downarrow\rangle$ and $|^{2}P_{3/2,+1/2};\uparrow\rangle$.\label{fig:StateCoeff}}
\end{figure*}

It is worth noting that, in the last decade, ultrashort laser pulses with the aforementioned characteristics have become largely available to experimentalists in the attosecond physics community. Moreover, since the entanglement between the parent ion and the photoelectron is encoded in the spin degree of freedom of the latter, and not in its kinetic energy, the time-duration of the photoionizing pulse is not a sensitive parameter for the formation of this type of entanglement. A short duration of the photoionizing pulse will result in an overlap in kinetic energy of the photoelectron wavepackets coming out of the ground-state spin-orbit doublet of the parent ion. On the contrary, a pulse with longer duration would give rise to two series of above-threshold-ionization peaks in the photoelectron spectrum, one shifted with respect to the other by the difference between the ionization potential of the ground and first-excited states of the argon cation. These two different scenarios should result in difference levels of entanglement (low and high, respectively) encoded in the photoelectron's kinetic energy, but we do not expect the entanglement encoded in the spin degree of freedom to be significantly affected.

\subsection{\label{subsec:results} Results and discussion.}

\begin{figure*}
\includegraphics[scale=1.0]{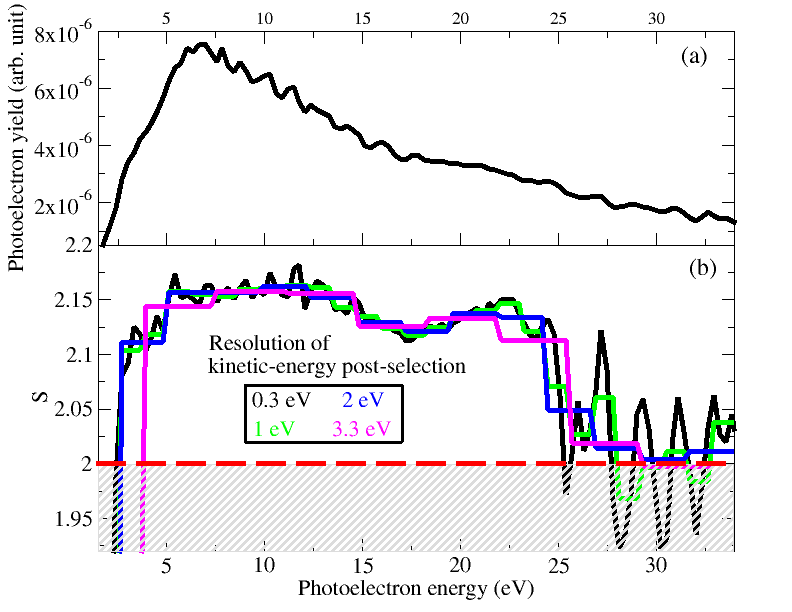}
\caption{Panel (a) - Angle-integrated, kinetic-energy dependent photoelectron yield upon photoionization of Ar by a circularly polarized 3-cycle laser pulse with peak intensity $I_{peak} = 5 \times 10^{13} W/cm^{2}$ and $\hbar\omega_{central}= 2$ eV central photon energy. 
Panel (b) - Kinetic-energy dependent optimized value of the Bell inequality quantity $\mathbb{S}$ from Eq.~\ref{eq:BellInequality}. Values greater than 2 correspond to a violation of the Bell inequality. Curves with different colors correspond to different resolutions in the post-selection of the value of the photoelectron's kinetic energy.\label{fig:BellViolation}}
\end{figure*}

In Fig.~\ref{fig:StateCoeff} we show the angle-integrated, kinetic energy-dependent matrix elements of the $\hat{\tilde{\rho}}^{\xi}$ density matrix of Eq.~\ref{eq:initial_matrix}, resulting from photoionization of argon atom by the laser pulse described in the previous subsection. The density matrix is normalized at each value of the photoelectron's kinetic energy. The diagonal matrix elements $\tilde{\rho}^{\xi}_{i,i}=P_{i}$ correspond to the populations $P_{i} $ of the 6 states of Eq.~\ref{eq:statetot} and Figs~(\ref{fig:Levels},\ref{fig:Scheme}). 
As it can be seen in panel (a) of Fig.~\ref{fig:StateCoeff}, our \textit{ab initio} numerical results confirm and quantify 
the strong predominance of ionization from the states $|^{2}P_{1/2,+1/2};\uparrow\rangle$, $|^{2}P_{3/2,+3/2};\downarrow\rangle$ and $|^{2}P_{3/2,+1/2};\uparrow\rangle$, here labelled states 1, 2 and 3, respectively, which reflects the strongly-favored emission of counter-rotating electrons (electrons whose orbital angular momentum has opposite sign with respect to the spin angular momentum of the ionizing photons) as it was predicted with elegant analytical techniques (yet applied to a more approximate model of the argon atom) in~\cite{Barth2013}. 
The off-diagonal matrix elements shown in Fig.~\ref{fig:StateCoeff} are the ones that contribute to the Bell inequality (see Eq.~\ref{Bell_explicit}): in the lower left panel, we show the real and imaginary parts of the angle-integrated, kinetic-energy dependent density matrix elements $\tilde{\rho}^{\xi}_{21}$ between the states $|^{2}P_{3/2,+3/2};\downarrow\rangle$ and $|^{2}P_{1/2,+1/2};\uparrow\rangle$, while in the lower right panel we show the real and imaginary parts of the $\tilde{\rho}^{\xi}_{43}$ matrix elements between the states $|^{2}P_{1/2,-1/2};\downarrow\rangle$ and $|^{2}P_{3/2,+1/2};\uparrow\rangle$. 

\begin{figure*}
\includegraphics[scale=1.0]{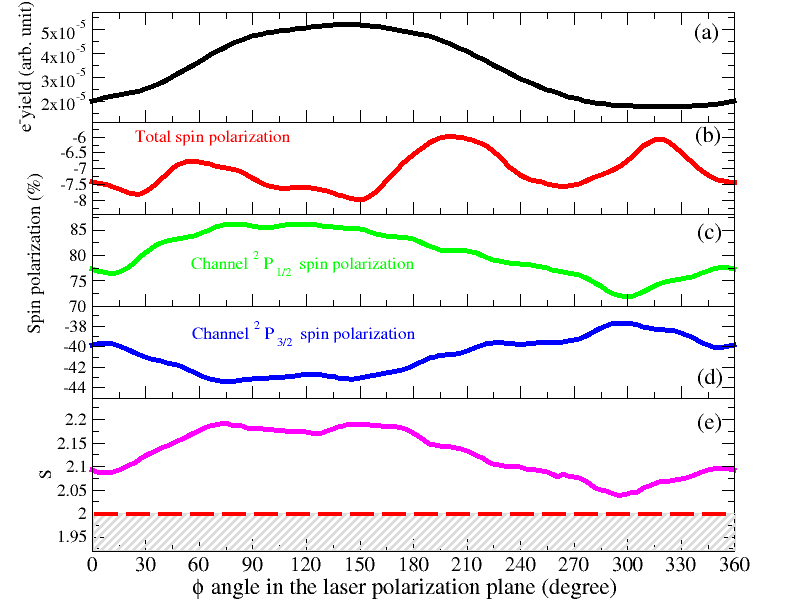}
\caption{Panel (a) - Kinetic energy-integrated, angle of emission dependent photoelectron yield upon photoionization of Ar by a circularly polarized 3-cycle laser pulse with peak intensity $I_{peak} = 5 \times 10^{13} W/cm^{2}$ and $\hbar\omega_{central}= 2$ eV central photon energy. 
Panel (b) - Angle of emission dependent, total spin polarization $Pol_{S}^{^{2}P_{1/2}}$ of the photoelectron emission (see Eq.~\ref{eq:polarization}). Panel (c) -  Same as (b) but for the spin polarization $P_{S}^{^{2}P_{1/2}}$ associated with the photoelectron emission from the first-excited state $|^{2}P_{1/2}\rangle$ channel of the parent ion (see Eq.~\ref{eq:polarization}).  Panel (d) - same as (c) but for the ground-state parent ion channel $|^{2}P_{3/2}\rangle$.
Panel (e) - Angle of emission dependent optimized value of the Bell inequality quantity $\mathbb{S}$ from Eq.~\ref{eq:BellInequality}. Values greater than 2 correspond to a violation of the Bell inequality. 
\label{fig:BellViolation_2}}
\end{figure*}

In Fig.~\ref{fig:BellViolation} we show the calculated photoelectron yield and the values of the Bell function $\mathbb{S}$ from Eq.~\ref{eq:BellInequality} obtained by post-selecting photoionization events characterized by a value of the photoelectron's kinetic energy in different ranges. Curves with different colors correspond to different resolutions in the post-selection of the value of the photoelectron's kinetic energy, from a minimum of 0.3 eV to a maximum of 3.3 eV. Here is it worth mentioning that the range of values over which performing the post-selection is only limited by the resolution of the photoelectron spectrometer and/or by the minimum/maximum kinetic energy detectable. The results of fig.~\ref{fig:BellViolation} show that the broad photoelectron yield extends over a wide range of kinetic energies, and it is peaked at around 7 eV. The broadness of the energy-dependent photoelectron spectrum, which lacks the peak structure typical of ATI, is due to the short duration of the laser pulse we used in this work. The laser parameters ($\frac{\Omega_{b}}{\Omega_{a}},\,\Omega_\text{eff} \times T,\,\phi$) that determine the ionic operators $\hat{Q}^{\text{Ion}}$ and $\hat{R}^{\text{Ion}}$ have been optimized, for each post-selection choice, in order to maximize the violation of the resulting Bell inequality. 
Values of $\mathbb{S}$ greater than 2 correspond to a violation of the Bell inequality and a direct detection of entanglement. Violation of the Bell inequality can be observed over a wide range of photoelectron kinetic energies. The maximum violation, with $\mathbb{S} \approx 2.17$ can be observed for events where the detected photoelectron has a kinetic energy around 11-12 eV. The ionic operators that construct, and maximize the violation of, the corresponding Bell inequality are characterized by the following parameters: in the case of the $\hat{Q}$ operator the ratio between the Rabi frequencies of the two applied resonant pulses is given by $\left(\frac{\Omega_{b}}{\Omega_{a}}\right)_{\hat{Q}}=2.458$, the duration of the laser pulses needs to satisfy $\left(\Omega_\text{eff} \times T \right)_{\hat{Q}} = 3\pi$, and the relative phase between the linear and circular polarized pulses is $\phi_{\hat{Q}} \approx 0$; in the case of the $\hat{R}$ operator the ratio between the Rabi frequencies of the two applied resonant pulses is given by $\left(\frac{\Omega_{b}}{\Omega_{a}}\right)_{\hat{R}}=0.41845$, the duration of the laser pulses needs to satisfy $\left(\Omega_\text{eff} \times T \right)_{\hat{R}} = 15\pi$, and the relative phase between the linear and circular polarized pulses also is $\phi_{\hat{R}} \approx 0$. 

In Fig.~\ref{fig:BellViolation_2} we show the kinetic energy-integrated photoelectron yield, the channel-resolved spin polarization for the $J=1/2$ and the $J=3/2$ ionization channels, and the values of the Bell function $\mathbb{S}$ from Eq.~\ref{eq:BellInequality} obtained by post-selecting photoionization events characterized by specific values of the photoelectron's emission angle $\Phi$ in the polarization plane of the photoionizing laser pulse.
The values of the calculated channel-resolved spin polarizations are found to be high and not too distant from the theoretical maxima of $100\%$  and $-50\%$ for the $^{2}P_{1/2}$ and $^{2}P_{3/2}$ channels respectively. 
The results show that disregarding the kinetic energy of the photoelectron and post-selecting events where emission happens in specific spatial directions gives rise to stronger violation of the resulting Bell inequalities with respect to the previously-presented case of post-selection over the value of the photoelectron's kinetic energy.

\begin{figure*}
\includegraphics[scale=0.75]{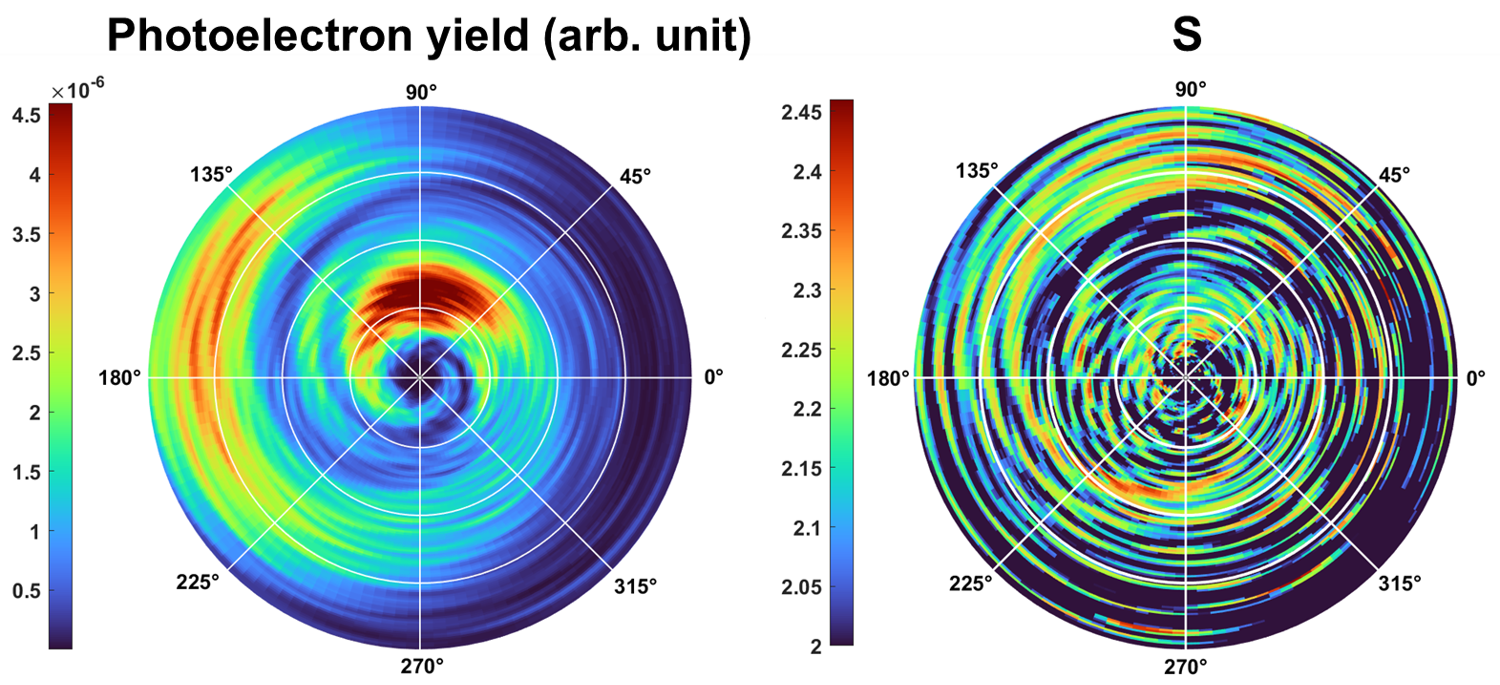}
\caption{Left plot - $\Phi$ Angle (in the laser polarization plane) and kinetic-energy dependent photoelectron yield upon photoionization of Ar by a circularly polarized 3-cycle laser pulse with peak intensity $I_{\text{peak}} = 5 \times 10^{13} W/cm^{2}$ and $\hbar\omega_{\text{central}}= 2$ eV central photon energy. 
The radial coordinate indicates the kinetic energy of the photoelectron; the interval between neighbor, concentric circles is 7 eV. 
Right plot - 
$\Phi$ Angle (in the laser polarization plane) and kinetic-energy dependent optimized value of the Bell inequality quantity $\mathbb{S}$ from Eq.~\ref{eq:BellInequality}. Values greater than 2 correspond to a violation of the Bell inequality. 
\label{fig:BellViolationAngle}}
\end{figure*}

In Fig.~\ref{fig:BellViolationAngle} we show the energy-dependent photoelectron angular distribution (PAD), and the values of the Bell function $\mathbb{S}$ from Eq.~\ref{eq:BellInequality} obtained by post-selecting photoionization events characterized both by specific values of the photoelectron's kinetic energy and of its emission angle $\Phi$ in the polarization plane of the photoionizing laser pulse. 
The results show that post-selecting events where, not only the emission happens in specific spatial directions, but also the photoelectron has specific kinetic energy, gives rise to the strongest violation of the resulting Bell inequalities, the value of $\mathbb{S}$ reaching $\approx 2.45$. 
The PAD is peaked around $\Phi_{1} = 90$ and $\Phi_{2} = 150$, at energies around $\xi_{1} = 9$ eV and $\xi_{2} = 23$ eV, respectively. At an angle $\Phi = 87$ and photoelectron energy $\xi = 8$ eV we have strong photoelectron signal and a robust violation given by $\mathbb{S} \approx 2.285$. The ionic operators that construct, and maximize the violation of, the corresponding Bell inequality are characterized by the following parameters: in the case of the $\hat{Q}$ operator the ratio between the Rabi frequencies of the two applied resonant pulses is given by $\left(\frac{\Omega_{b}}{\Omega_{a}}\right)_{\hat{Q}}=2.4391$, the duration of the laser pulses needs to satisfy $\left(\Omega_\text{eff} \times T \right)_{\hat{Q}} = 3\pi$, and the relative phase between the linear and circular polarized pulses is $\phi_{\hat{Q}} \approx 0$; in the case of the $\hat{R}$ operator the ratio between the Rabi frequencies of the two applied resonant pulses is given by $\left(\frac{\Omega_{b}}{\Omega_{a}}\right)_{\hat{R}}=0.40121$, the duration of the laser pulses needs to satisfy $\left(\Omega_\text{eff} \times T \right)_{\hat{R}} = 19\pi$, and the relative phase between the linear and circular polarized pulses also is $\phi_{\hat{R}} \approx 0$. Here it is interesting to note that these values are very similar to the ones that maximize the energy-dependent only Bell inequality. 

Our results show that high values of the channel-resolved spin polarizations favor the possibility of strong violation of the Bell inequality we constructed in this work. The relationship between the value of the channel-resolved spin polarization and the strength of the violation is evident from Fig.~\ref{fig:BellViolation_2}, where one can notice how peaks in the former quantity correspond to peaks in the corresponding value of $\mathbb{S}$. This is also confirmed by the lack of violation of the Bell inequality of Eq.~\ref{eq:BellInequality} when the populations of the states 
$|^{2}P_{1/2,-1/2};\;\xi,\,\bf{\downarrow}\rangle$, 
$|^{2}P_{3/2,-1/2};\;\xi,\,\bf{\downarrow}\rangle$ and
$|^{2}P_{3/2,-3/2};\;\xi,\,\bf{\uparrow}\rangle$ are artificially increased by a factor of 8. Indeed, the artificial increase of these populations corresponds to artificial increase of the photoionization yield from the co-rotating $p_{+1}$ atomic orbital of Ar and gives rise to an artificial suppression of the spin polarization corresponding to each ionization channel.

\section{\label{sec:lev5}Conclusion and outlook}

The results presented here predict that strong violation of Bell inequalities can be observed in 
standard ultrafast photoionization experiments. This challenges the appropriateness of treating the 
photoelectron and the ion as individual systems, and highlights the experimental relevance of 
electron-ion correlations.

The present work paves the way to the direct observation of entanglement in the context of ultrafast 
photoionization of many-electron systems, the latter not being restricted to the case of 
strong-field ionization of atomic systems. In fact, first, the entanglement protocol we constructed 
in this work could be extended to other systems where quantum entanglement can be encoded in the 
spin degree of freedom of the spin-polarized photoelectron such as, for example, linear and ring-
shaped molecules with a degenerate HOMO and a singlet ground state~\cite{Barth2013}. In such systems 
spin-orbit states have lower degeneracy than in atoms and higher degree of total integrated spin 
polarization are expected. Moreover, other photoionization regimes can feature in spin polarization 
of the photoelectron beam such as, for example, single-photon ionization, either from a particular 
fine structure level of an atom or a molecule~\cite{Cherepkov1981} or in the vicinity of the Cooper 
minima in the photoionization continua~\cite{Fano1969}, and resonant multi-photon ionization in the 
perturbative limit~\cite{Lambropoulos1973,Dixit1981}. Moreover, it is worth noting that the 
detection of entanglement in attosecond photoionization is not necessarily limited to the 
measurement of spin degree of freedom of the photoelectron, but can in general also be extended to 
Bell tests based on the observation of the photoelectron's orbital angular momentum that is already 
becoming experimentally accessible~\cite{Grillo2017a,Grillo2017b,Larocque2018}. 

Our work opens up a novel perspective on attosecond physics, directed towards the development of new 
types of spectroscopy targeted to detect the different types of quantum correlation present in 
atto-ionized states of quantum systems. A key step forward for future work on the subject consists 
of extending this type of protocols to molecular systems and, in particular, investigating the 
quantum correlations that can exist between the state of the photoelectron and the outcome of a 
chemical reaction pathway in the parent, excited molecular ion, with the ultimate goal of giving an 
answer to the following research question: what is the dependence of the observed hole migration 
dynamics in the molecular ion on the detection scheme of the entangled state created by 
photoionization? 

Finally, our work also opens up the possibility to coherently control entanglement in photoionized 
systems by performing photoionizing pulse shaping with the aim of detecting specific quantum 
correlations in coincidence measurements.

\begin{acknowledgments}
The authors acknowledge support from the EPSRC grant "\textit{Quantum entanglement in attosecond ionization}", grant number EP/V009192/1.
\end{acknowledgments}

\appendix

\section{\label{sec:app1}Effective reduced density matrix.}

The final state of the photoionized atom can be written as
\begin{subequations}
\label{eq:FullState}
\begin{equation}
|\Psi^{F}_{N}\rangle =  \sum_{j}\sum_{\bf{s}}\iint d\xi d\alpha\, C_{j}\left(\xi,\bf{s},\alpha \right) |j;\;\xi,\,\bf{s},\,\alpha \rangle,
\label{eq:state1}
\end{equation}
\begin{equation}
|j;\;\xi,\,\bf{s},\,\alpha \rangle\,=\,
|I^{+}_{j}\rangle \, \otimes \,|e^{-}_{\xi,\,\bf{s},\,\alpha}\rangle ~.
  \label{eq:state2}  
\end{equation}
\end{subequations}

In Eqs.~(\ref{eq:FullState}), the index j labels the different internal bound states of the parent ion, $\xi$ is the kinetic energy of the photoelectron, $\bf{s}$ its spin component along a specific axis in space (in the present case the propagation direction of the photoionizing laser beam) and $\alpha$ denotes any extra continuous degree of freedom describing the state of the photoelectron, e.g. the emission angles. 

The density matrix of the full composite system can be written as 
\begin{equation}
\rho_{i\xi \alpha \bf{s},\,j \xi^{'} \alpha^{'} \bf{s^{'}}}\,=\,
C_{i}\left(\xi,\bf{s},\alpha \right)\,\cdot
C_{j}^{*}\left(\xi^{'},\bf{s^{'}},\alpha^{'} \right) 
\label{eq:density1}
\end{equation}

Within a specific measurement protocol, it is possible to either ignore (i.e. do not measure) some of the observables characterizing a particular subsystem or to use these observables to post-select the events in which only some specific outcomes of their measurement are observed. 
The former scenario is equivalent to considering an effective reduce density matrix, which can be obtained from Eqs.~(\ref{eq:density1}) by tracing over the unobserved degree of freedom, e.g. in the case of $\alpha$

\begin{eqnarray}
&\hat{\tilde{\rho}}\,=\, \textit{Tr}_{\alpha}\left(\hat{\rho}\right),&\label{eq:density2} \\
&\tilde{\rho}_{i\xi \bf{s},\,j \xi^{'} \bf{s^{'}}}\,=\,
\int d\alpha \, C_{i}\left(\xi,\bf{s},\alpha \right) \times
C_{j}^{*}\left(\xi^{'},\bf{s^{'}},\alpha \right).&\label{eq:density3}   
\end{eqnarray}

Post-selection over the value of the photoelectron's kinetic energy gives rise to a final effective density matrix parametrized with respect to $\xi$: 
\begin{equation}
\tilde{\rho}^{\xi}_{i \bf{s},\,j\bf{s^{'}}}\,=\,
\int d\alpha \, C_{i}\left(\xi,\bf{s},\alpha \right) \times
C_{j}^{*}\left(\xi,\bf{s^{'}},\alpha \right).
\label{eq:density4} 
\end{equation}
In the case where the degree of freedom $\alpha$ consists of the direction of emission of the photoelectron, as it can be measured in a photoelectron angular distribution experiment, we have for example:
\begin{equation}
\tilde{\rho}^{\xi}_{i \bf{\uparrow},\,j\bf{\downarrow}}\,=\,
\int \, C_{i,\uparrow}\left(\xi,\theta,\phi \right) \times
C_{j,\downarrow}^{*}\left(\xi,\theta,\phi \right)
  \sin\left(\theta \right) d\theta d\phi.
\label{eq:densityANGLE} 
\end{equation}
The diagonal elements of this effective density matrix are given by 
\begin{equation}
\tilde{\rho}^{\xi}_{i \bf{s},\,i\bf{s}}\,=\,
\int d\alpha \, |C_{i}\left(\xi,\bf{s},\alpha \right)|^{2}\,=\, |\tilde{C}_{i}\left(\xi,\bf{s}\right)|^{2} .
\label{eq:densityDIAG} 
\end{equation}
Here it is important to note that the off-diagonal matrix elements $\tilde{\rho}^{\xi}_{i \bf{s},\,j\bf{s^{'}}}$ cannot be in general simplified by the real-valued product $|\tilde{C}_{i}\left(\xi,\bf{s}\right)| \times |\tilde{C}_{j}\left(\xi,\bf{s^{'}}\right)|$, rather they read as
\begin{eqnarray}
    \tilde{\rho}^{\xi}_{i \bf{s},\,j\bf{s^{'}}}\,=\, 
     |\tilde{C}_{i}\left(\xi,\bf{s}\right)|  |\tilde{C}_{j}\left(\xi,\bf{s^{'}}\right)|
     \times \nonumber \\ 
     \times \, 
     G_{i \bf{s},\,j\bf{s^{'}}}\left(\xi \right) \times \textit{e}^{\textit{i} \tilde{\chi}^{\xi}_{i\bf{s},\,j\bf{s^{'}}}} ~.
     \label{eq:densityOFFDIAG} 
\end{eqnarray}
In Eq.~\ref{eq:densityOFFDIAG}, 
\[G_{i \bf{s},\,j \bf{s^{'}}} (\xi)\,=\, 
\frac{|\tilde{\rho}^{\xi}_{i \bf{s},\,j\bf{s^{'}}}|}
{\sqrt{\tilde{\rho}^{\xi}_{i \bf{s},\,i\bf{s}}\times\tilde{\rho}^{\xi}_{j \bf{s^{'}},\,j\bf{s^{'}}}}}\]
is the energy-dependent degree of coherence between the 
$i, \bf{s}$ and the $j, \bf{s^{'}}$ states of the composite parent ion and photoelectron system.

\section{\label{sec:app2}Derivation of the Bell inequality}
The pulse scheme introduced in Sec.~\ref{sec:lev2} does not affect the populations of states $^{2}P_{3/2,-3/2}$ and $^{2}P_{3/2,-1/2}$. Therefore, it is possible to discard these states in the calculation of the Bell inequality and work within a smaller Hilbert space. 
The effective density matrix of the Ar$^{+}$ parent ion produced by the photoionization process reads as
\begin{equation}
\tilde{\rho}^{\xi}= \left(
    \begin{array}{cccc}
|\tilde{C}_{1,\uparrow}\left(\xi\right)|^{2}
& \tilde{\rho}_{1 \bf{\uparrow},2\bf{\downarrow}}\left(\xi \right) & \tilde{\rho}_{1 \bf{\uparrow},3\bf{\uparrow}}\left(\xi \right) & \tilde{\rho}_{1 \bf{\uparrow},4\bf{\downarrow}}\left(\xi \right) \\
\tilde{\rho}^{*}_{1 \bf{\uparrow},2\bf{\downarrow}}\left(\xi \right) & |\tilde{C}_{2,\downarrow}\left(\xi\right)|^{2} & \tilde{\rho}_{2 \bf{\downarrow},3\bf{\uparrow}}\left(\xi \right) & \tilde{\rho}_{2 \bf{\downarrow},4\bf{\downarrow}}\left(\xi \right) \\
\tilde{\rho}^{*}_{1 \bf{\uparrow},3\bf{\uparrow}}\left(\xi \right) & \tilde{\rho}^{*}_{2 \bf{\downarrow},\,3\bf{\uparrow}}\left(\xi \right) & |\tilde{C}_{3,\uparrow}\left(\xi\right)|^{2} & \tilde{\rho}_{3 \bf{\uparrow},4\bf{\downarrow}}\left(\xi \right) \\
\tilde{\rho}^{*}_{1 \bf{\uparrow},4\bf{\downarrow}}\left(\xi \right) & \tilde{\rho}^{*}_{2 \bf{\downarrow},4\bf{\downarrow}}\left(\xi \right) & \tilde{\rho}^{*}_{3 \bf{\uparrow},4\bf{\downarrow}}\left(\xi \right) & |\tilde{C}_{4,\downarrow}\left(\xi\right)|^{2} \\
\end{array}
\right)
\label{eq:initial_matrix}
\end{equation}
The $|^{2}S\rangle$ bound excited state is not populated by the photoionization process and it has therefore been omitted in Eq.~\ref{eq:initial_matrix}.

The measurement protocol on the parent ion subsystem consists of: 
\begin{itemize}
    \item Application of the two synchronized resonant laser pulses as described in Fig.~\ref{fig:Scheme}.
    \item Detecting the fluorescence from the $^{2}S$ excited state.
\end{itemize}
Measuring the fluorescence emission resulting from the population of this upper state corresponds to the following operator $\hat{O}^{\text{Ion}}$, expressed in the basis of the parent ion energy eigenstates as:
\begin{equation}
    \hat{O}^{\text{Ion}} \,=\, \left( 
    \begin{array}{cccccc}
         -1 &  0 & 0 & & & \\
        0 & -1 & 0 & & 0 & \\
        0 & 0 & +1 & & & \\
          &  &  & -1 & 0 & 0 \\
          &  0 &  & 0 & -1 & 0 \\
          &  &  & 0 & 0 & +1
    \end{array}
    \right)
\end{equation}

The effective operator for the ionic measurement can be written as 
\begin{eqnarray}
    \hat{Q}^{\text{Ion}} \,=\, \hat{U}^{\dagger} \hat{O}^{\text{Ion}} \hat{U} \\
    \hat{U} \,=\, \textit{e}^{-\textit{i}\hat{H}^{\text{I}}\,T}
\end{eqnarray}
where $\hat{U}$ is the unitary propagator constructed with the following interaction Hamiltonian, in the interaction picture, 
\begin{equation}
    \hat{H}^{\text{I}} =\frac{1}{2} \left( 
    \begin{array}{cccccc}
        0 & 0 & \Omega_{a} & & & \\
        0 & 0 & \Omega_{b}\textit{e}^{-\textit{i}\phi} & & 0 & \\
        \Omega_{a} & \Omega_{b}\textit{e}^{\textit{i}\phi}  & 0 & & & \\
          &  &  & 0 & 0 & \frac{\Omega_{b}}{\sqrt{3}}\textit{e}^{-\textit{i}\phi} \\
          &  0 &  & 0 & 0 & - \Omega_{a} \\
          &  &  & \frac{\Omega_{b}}{\sqrt{3}}\textit{e}^{\textit{i}\phi} &
          -\Omega_{a} &
          0
    \end{array}
    \right) ~.
    \label{eq:ham_int}
\end{equation}
In Eq.~\ref{eq:ham_int}, $\Omega_{a}=E_{\text{Lin}}d_{A}$ and $\Omega_{a}=E_{\text{Cir}}d_{B}$ are the Rabi frequencies corresponding to the linear and circularly polarized fields, respectively, and $\phi$ is the relative phase between the two applied laser fields.
This corresponds to two decoupled 3-level $\Lambda$ systems. 
Diagonalization of the interaction Hamiltonian $\hat{H}^{\text{I}}$ allows one to obtain a general form for the effective measurement operator on the parent ion subsystem, expressed as
\begin{subequations}
\label{eq:Ionic_operator}
\begin{equation}
    \hat{Q}^{\text{Ion}} \,=\,\left( 
    \begin{array}{cccc}
       Q_{11} & Q_{12} & 0 & 0 \\
       Q_{12}^{*} & Q_{22} & 0 & 0 \\
       0 & 0 & Q_{33} & Q_{34} \\
       0 & 0 & Q_{34}^{*} & Q_{44}
    \end{array}
    \right)\,,
    \end{equation}
where the matrix elements in the upper-left block read as
    \begin{equation}
        Q_{11} \,=\, \frac{1}{\Omega_\text{eff}^{2}}\left\{ \Omega_{a}^{2}\left[ \sin^{2}\left(\frac{\sigma}{2}\right)\,-\,\cos^{2}\left(\frac{\sigma}{2}\right)\right] - \Omega_{b}^{2}\right\}\,,
    \end{equation}
    \begin{equation}
        Q_{22} \,=\, \frac{1}{\Omega_\text{eff}^{2}}\left\{ \Omega_{b}^{2}\left[ \sin^{2}\left(\frac{\sigma}{2}\right)\,-\,\cos^{2}\left(\frac{\sigma}{2}\right)\right] - \Omega_{a}^{2}\right\}\,,
    \end{equation}
    \begin{equation}
        Q_{12} \,=\, \frac{2\Omega_{a}\Omega_{b}}{\Omega_\text{eff}^{2}} \sin^{2}\left(\frac{\sigma}{2}\right)\,\textit{e}^{+\textit{i}\phi}\,,
    \end{equation}
and the $\Omega_{\text{eff}}$ and $\sigma$ parameters are given by
    \begin{equation} 
\Omega_\text{eff}\,=\,\sqrt{\Omega_{a}^{2}+\Omega_{b}^{2}};\,\,\,\,\sigma\,=\,\Omega_\text{eff}\,T ~.
    \end{equation}
The matrix elements in the lower-right block ($Q_{33}$, $Q_{34}$ and $Q_{44}$) can be obtained from the correspondent ones in the upper-left block as
   
\begin{eqnarray}   
        Q_{33}\left(\Omega_{a},\Omega_{b}\right)\,=\,Q_{22}\left(-\Omega_{a},\frac{\Omega_{b}}{\sqrt{3}}\right)  \nonumber \\ 
      Q_{44}\left(\Omega_{a},\Omega_{b}\right)\,=\,Q_{11}\left(-\Omega_{a},\frac{\Omega_{b}}{\sqrt{3}}\right)  \nonumber \\ 
      Q_{34}\left(\Omega_{a},\Omega_{b}\right)\,=\,Q_{12}^{*}\left(-\Omega_{a},\frac{\Omega_{b}}{\sqrt{3}}\right)
    \end{eqnarray}
\end{subequations}
Here we have included only the matrix elements of the $\hat{Q}$ operator on the Hilbert space of the parent ion that is initially populated by the ionizing pulse, i.e. excluding the $|^{2}S\rangle$ doublet.

Using this type of measurement for the parent ion subsystem, combined with the spin measurements for the photoelectron, we can write the following Bell inequality:  
\begin{eqnarray}
    \mathbb{S} \,=\, \langle \hat{Q}^{\text{Ion}}\otimes2\hat{S}_{x}\rangle\,-\,\langle\hat{Q}^{\text{Ion}}\otimes2\hat{S}_{z}\rangle\,+\, \nonumber \\ +\,\langle\hat{R}^{\text{Ion}}\otimes2\hat{S}_{x}\rangle\,+\,\langle\hat{R}^{\text{Ion}}\otimes2\hat{S}_{z}\rangle\,.
    \label{eq:Bell}
\end{eqnarray}
The second ionic operator $\hat{R}^{\text{Ion}}$ has the same structure of $\hat{Q}^{\text{Ion}}$,  but it is characterized by different values of the laser parameters.
Since both the ionic and the photoelectron operators defined above have eigenvalues equal to $\pm 1$, the condition for detection of quantum entanglement between the photoelectron and the parent ion reads as
\begin{equation}
    \mathbb{S} \,\geq\, 2 ~.
\end{equation}
The operators acting on the composite bipartite system read as
\begin{equation}
    \hat{Q}^{\text{Ion}}\otimes2\hat{S}_{z} \,=\,\left( 
    \begin{array}{cccc}
       +Q_{11} & 0 & 0 & 0 \\
       0 & -Q_{22} & 0 & 0 \\
       0 & 0 & +Q_{33} & 0 \\
       0 & 0 & 0 & -Q_{44}
    \end{array}
    \right)\,,
    \end{equation}
and
\begin{equation}
    \hat{Q}^{\text{Ion}}\otimes2\hat{S}_{x} \,=\,\left( 
    \begin{array}{cccc}
       0 & Q_{12} & 0 & 0 \\
       Q_{12}^{*} & 0 & 0 & 0 \\
       0 & 0 & 0 & Q_{34} \\
       0 & 0 & Q_{34}^{*} & 0
    \end{array}
    \right) ~.
    \end{equation}
Using the density matrix of Eq.~\ref{eq:initial_matrix} to compute the traces, we finally obtain:    
\begin{widetext}
\begin{eqnarray} 
    \mathbb{S} \,=\, +\,R_{11}|\tilde{C}_{1,\uparrow}\left(\xi\right)|^{2}\,-\,R_{22}|\tilde{C}_{2,\downarrow}\left(\xi\right)|^{2}\,+\,R_{33}|\tilde{C}_{3,\uparrow}\left(\xi\right)|^{2} \,-\,R_{44}|\tilde{C}_{4,\downarrow}\left(\xi\right)|^{2} \nonumber \\ \,-\,
    \left(\,Q_{11}|\tilde{C}_{1,\uparrow}\left(\xi\right)|^{2}\,-\,Q_{22}|\tilde{C}_{2,\downarrow}\left(\xi\right)|^{2}\,+\,Q_{33}|\tilde{C}_{3,\uparrow}\left(\xi\right)|^{2} \,-\,Q_{44}|\tilde{C}_{4,\downarrow}\left(\xi\right)|^{2} \right) \,+\, \nonumber \\
\,+\,2Re\left[Q_{12}\tilde{\rho}^{*}_{1 \bf{\uparrow},\,2\bf{\downarrow}}\left(\xi \right)
\,+\,R_{12}\tilde{\rho}^{*}_{1 \bf{\uparrow},\,2\bf{\downarrow}}\left(\xi \right)
\,+\,Q_{34}\tilde{\rho}^{*}_{3 \bf{\uparrow},\,4\bf{\downarrow}}\left(\xi \right)
\,+\,R_{34}\tilde{\rho}^{*}_{3 \bf{\uparrow},\,4\bf{\downarrow}}\left(\xi \right)\right] ~.
\label{Bell_explicit}
\end{eqnarray}   
\end{widetext}

\nocite{*}

\bibliography{Main_Article}

\end{document}